\documentclass{article}
\usepackage[T1]{fontenc} 
\usepackage[utf8]{inputenc} 
\usepackage{ismir,amsmath,amsfonts,cite,url}
\usepackage{graphicx}
\usepackage{color}
\usepackage{array}
\usepackage{mathtools}
\usepackage[algo2e]{algorithm2e} 
\usepackage{algorithm}
\usepackage{booktabs}
\usepackage{multirow,tabularx}
\usepackage{dblfloatfix}

\DeclareMathSymbol{\shortminus}{\mathbin}{AMSa}{"39}
\everymath{\displaystyle}

\usepackage{tikz}
\usetikzlibrary{shapes.geometric, arrows}

\title{Adversarial Learning for Improved Onsets and Frames Music Transcription}

\oneauthor{Jong Wook Kim, Juan Pablo Bello}{Music and Audio Research Lab, New York University\\{\tt \{jongwook,jpbello\}@nyu.edu}}

\renewcommand{\L}{\mathcal{L}}
\newcommand{\E}{\mathbb{E}}
\newcommand{\R}{\mathbb{R}}
\newcommand{\x}{\mathbf{x}}
\newcommand{\y}{\mathbf{y}}
\newcommand{\z}{\mathbf{z}}
\newcommand{\X}{\mathbf{X}}
\newcommand{\Y}{\mathbf{Y}}

\renewcommand\thetable{\arabic{table}}
\begin{document}

\maketitle
\begin{abstract}
Automatic music transcription is considered to be one of the hardest problems in music information retrieval, yet recent deep learning approaches have achieved substantial improvements on transcription performance.
These approaches commonly employ supervised learning models that predict various time-frequency representations, by minimizing element-wise losses such as the cross entropy function.
However, applying the loss in this manner assumes conditional independence of each label given the input, and thus cannot accurately express inter-label dependencies.
To address this issue, we introduce an adversarial training scheme that operates directly on the time-frequency representations and makes the output distribution closer to the ground-truth.
Through adversarial learning, we achieve a consistent improvement in both frame-level and note-level metrics over Onsets and Frames, a state-of-the-art music transcription model.
Our results show that adversarial learning can significantly reduce the error rate while increasing the confidence of the model estimations.
Our approach is generic and applicable to any transcription model based on multi-label predictions, which are very common in music signal analysis.
\end{abstract}
\section{Introduction}\label{sec:introduction}

Automatic music transcription (AMT) concerns automated methods for converting acoustic music signals into some form of musical notation~\cite{benetos2013amt}.
AMT is a multifaceted problem and comprises a number of subtasks, including multi-pitch estimation (MPE), note tracking, instrument recognition, rhythm analysis, score typesetting, etc.
MPE predicts a set of concurrent pitches that are present at each instant, and it is closely related to the task of note tracking, which predicts the onset and offset timings of every note in audio.
In this paper, we address an issue in the recent approaches for MPE and note tracking, where the probabilistic dependencies between the labels are often overlooked.

A common approach for MPE and note tracking is through the prediction of a two-dimensional representation that is defined along the time and frequency axes and contains the pitch tracks of notes over time.
Piano rolls 
are the most common example of such representations, and deep salience~\cite{bittner2017deepsalience} 
is another example that can contain more granular information on pitch contours.
Once such representation is obtained, pitches and notes can be decoded by thresholding~\cite{kelz2016framewise} or other heuristic methods~\cite{kim2018crepe,hawthorne2018onsetsframes}.

To train a model that predicts a two-dimensional target representation $\hat{\Y} \in \R^{P \times T}$ from an input audio representation $\X$, where $P$ is the number of pitch labels and $T$ is the number of time frames, a common approach is to minimize the element-wise sum of a loss function $\L$:
\begin{equation}\label{eqn:elementwise}
\small
\textrm{minimize} ~~ \L(\hat{\Y}, \Y) = \sum_{p=1}^{P}  \sum_{t=1}^{T} \L (\hat{\Y}_{pt}, \Y_{pt}),
\end{equation}
where $\Y \in \mathbb{R}^{P \times T}$ is the ground truth.
In a probabilistic perspective, we can interpret $\L$ as the negative log-likelihood of the model parameters $\vartheta$ of a discriminative model $p_\vartheta(\Y | \X)$:
\begin{equation}
\small\thickmuskip=-0.1\thickmuskip
p_\vartheta(\Y | \X) ~=~ e^{\shortminus\L(\hat{\Y}\hspace{-0.05em}, \Y)} = \prod_{p=1}^{P} \prod_{t=1}^{T} e^{\shortminus\L(\hat{\Y}_{\hspace{-0.2em}pt}\hspace{-0.05em}, \Y_{\hspace{-0.2em}pt})} = \prod_{p=1}^{P} \prod_{t=1}^{T} p_\vartheta(\Y_{\hspace{-0.2em}pt} | \X)
\end{equation}
which indicates that each element of the label $\Y$ is conditionally independent with each other given the input $\X$.
This encourages the model to predict the average of the posterior, making blurry predictions when the posterior distribution is multimodal, e.g. natural images~\cite{dosovitskiy2016generating}.

Music data is highly contextual and multimodal, and the conditional independence assumption does not hold in general.
This is why many computational music analysis models employ a separate post-processing stage after sequence prediction.
One approach is to factorize the joint probability using the chain rule and assume the Markov property:
\begin{equation}
\small
p_\vartheta(\Y|\X) \approx \prod_{p=1}^{P} \prod_{t=1}^{T} p_\vartheta(\Y_{pt} | \Y_{\cdot(t-1)}, \X).
\end{equation}
This corresponds to appending hidden Markov models (HMMs) \cite{poliner2006discriminative} or recurrent neural networks (RNNs) \cite{sigtia2016endtoend,hawthorne2018onsetsframes} to the transcription model.
The Markov assumption is effective for one-dimensional sequence prediction tasks, such as chord estimation~\cite{ni2012hmm} and monophonic pitch tracking~\cite{mauch2014pyin}, but when predicting a two-dimensional representation, it still does not address the inter-label dependencies along the frequency axis.

There exist a number of models in the computer vision literature that can express inter-label dependencies in two-dimensional predictions, such as the neural autoregressive distribution estimator (NADE)~\cite{larochelle2011nade}, PixelRNN~\cite{van2016pixelrnn}, and PixelCNN~\cite{van2016pixelcnn}.
However, apart from a notable exception using a hybrid RNN-NADE approach~\cite{boulanger2012rnnnade}, the effect of learning the joint posterior distribution for polyphonic music transcription has not been well studied.

To this end, we propose a new approach for effectively leveraging inter-label dependencies in polyphonic music transcription. We pose the problem as an image translation task and apply an adversarial loss incurred by a discriminator network attached to the baseline model.
We show that our approach can consistently and significantly reduce the transcription errors in \emph{Onsets and Frames} \cite{hawthorne2018onsetsframes}, a state-of-the-art music transcription model.

\section{Background}

\subsection{Automatic Transcription of Polyphonic Music}

Automatic transcription models for polyphonic music can be classified into frame- or note-level approaches.
Frame-level transcription is synonymous with multi-pitch estimation (MPE) and operates on tiny temporal slices of audio, or frames, to predict all pitch values present in each frame.
Note-level transcription, or note tracking, operates at a higher level, predicting a sequence of note events that contains the pitch, the onset time, and optionally the offset time of each note.
Note tracking is typically implemented as a post-processing stage on the output of MPE~\cite{benetos2019amt}, by connecting and grouping the pitch estimates over time to produce discrete note events.
In this sense, we can say that MPE is at the core of polyphonic music transcription.

Two categories of approaches for MPE have been most successful in recent years: matrix factorization and deep learning.
Factorization-based models for music transcription~\cite{smaragdis2003nmf} use non-negative matrix factorization (NMF)~\cite{lee2001nmf} to factorize a time-frequency representation $\X \in \R^{F \times T}$ as a product of a dictionary matrix $\mathbf{D} \in \R^{F \times K}$ and an activation matrix $\mathbf{A} \in \R^{K \times T}$, where $K$ is the number of pitch labels to be transcribed, e.g. 88 keys for piano transcription.
This allows for an intuitive interpretation of each matrix, where each column of $\mathbf{D}$ contains a spectral template for a pitch label, and each row of $\mathbf{A}$ contains the activation of the corresponding pitch over time.
Various extensions of factorization-based methods 
have been proposed to leverage sparsity~\cite{abdallah2006sparse}, adaptive estimation of harmonic spectra~\cite{vincent2010adaptive,fuentes2013harmonic}, and modeling of attack and decay sounds~\cite{benetos2013multiple,ewert2016admm}.
In all of these approaches, an iterative gradient-descent algorithm is used to minimize an element-wise divergence function between the matrix factorization $\mathbf{DA}$ and the target matrix $\mathbf{X}$~\cite{fevotte2011betanmf}.

Deep learning~\cite{lecun2015deeplearning} methods for music transcription are increasingly popular
~\cite{benetos2019amt}, as larger labeled datasets and more powerful hardware become accessible.
These approaches use neural networks (NN) to produce music transcriptions from the input audio.
An early work~\cite{nam2011classification} used deep belief networks~\cite{hinton2006dbn} to extract audio features which are subsequently fed to pitch-wise SVM-HMM pairs to predict the target piano rolls.
More recent approaches are based on convolutional
~\cite{bittner2017deepsalience,kelz2016framewise} and/or recurrent neural networks~\cite{bock2012rnn,sigtia2016endtoend,hawthorne2018onsetsframes}, which are also optimized with gradient descent to minimize an element-wise loss of predicting the target time-frequency representations.

\begin{figure}[t]
	\centering
	\includegraphics[width=\columnwidth]{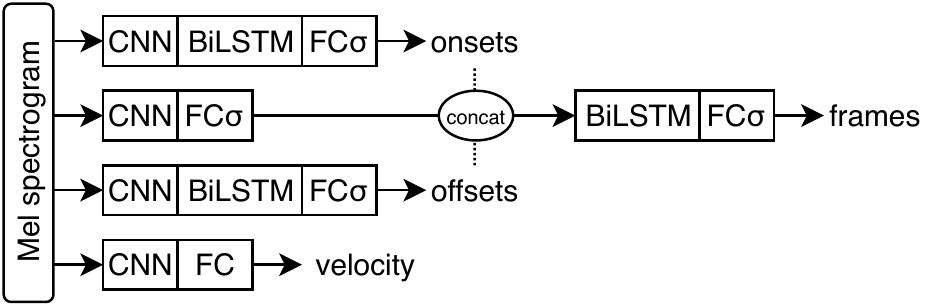}
	\caption{The Onset and Frames model. CNN denotes the convolutional acoustic model taken from~\cite{kelz2016framewise}, FC denotes a fully connected layer, and $\sigma$ denotes sigmoid activation. Dotted lines mean stop-gradient, i.e. no backpropagation.}\label{fig:onsetsframes}
\end{figure}

Onsets and Frames~\cite{hawthorne2018onsetsframes} is a state-of-the-art piano transcription model that we use as our baseline.
It uses multiple columns of convolutional and recurrent layers to predict onsets, offsets, velocities, and frame labels from the Mel spectrogram input, as shown in Figure~\ref{fig:onsetsframes}.
Predicted onset and frame posteriors are then used for decoding the note sequences, where a threshold value is used to create binary onset and frame activations, and frame activations without the corresponding onsets are disregarded.

As discussed above, most NMF- and NN-based methods, including Onsets and Frames, use an element-wise optimization objective which does not consider the inter-label dependencies.
This motivates the adversarial training scheme that is outlined in the following subsection.

\subsection{Generative Adversarial Networks and \texttt{pix2pix}}

Generative adversarial networks (GANs)~\cite{goodfellow2014gan} refer to a family of deep generative models which consist of two components, namely the generator $G$ and the discriminator $D$.
Given a data distribution $\x \sim p(\x)$ and latent codes $\z \sim p(\z)$, GAN performs the following minimax game:
\begin{equation}\label{eqn:gan}
\small
\min_{G} \max_{D} \underbrace{\Big [ \E_{\x} \log D(\x) + \E_{\z} \log (1 - D ( G(\z))) \Big ]}_{\L_{\text{GAN}}(G, D)}.
\end{equation}
$G$ and $D$ are implemented as neural networks trained in an adversarial manner, where the discriminator learns to distinguish the generated samples from the real data, while the generator learns to produce realistic samples to fool the discriminator.
GANs are most renowned for their ability to produce photorealistic images~\cite{karras2018stylegan} and have shown promising results on music generation as well~\cite{engel2019gansynth,dong2018musegan,yang2017midinet}.
We refer the readers to~\cite{goodfellow2016gan,creswell2018gan} for a comprehensive review of the techniques, variants, and applications of GANs.

The second term in Equation~\ref{eqn:gan} has near-zero gradients when $D(G(\z)) \approx 0$, which is usually the case in early training. To avoid this, a non-saturating variant of GAN is suggested in~\cite{goodfellow2014gan} where the generator is trained with the following optimization objective instead:
\begin{equation}\label{eqn:nsgan}
\small
\max_{G}~ \E_{\z} \log D(G(\z)).
\end{equation}
The non-saturating GAN loss is used more often than the minimax loss in Equation \ref{eqn:gan} and is implemented by flipping the labels of fake data while using the same loss function.
Least-squares GAN \cite{mao2017lsgan} is an alternative method to address the vanishing gradient problem, which replaces the cross entropy loss in Equations \ref{eqn:gan}-\ref{eqn:nsgan} with squared errors:%
{\def\arraystretch{2}\begin{equation}\label{eqn:lsgan}\small
\begin{array}{l@{}l}
\min_{D} ~\Big [ \E_{\x} (D(\x) - 1)^2 + \E_{\z} D(G(\x))^2 \Big ], \\
\min_{G} ~\>\E_{\z} (D(G(\z)) - 1)^2.
\end{array}
\end{equation}
}

While the default formulation of GAN concerns unconditional generation of samples from $p(\mathbf{x})$, conditional GANs (cGAN)~\cite{mirza2014cgan} produce samples from a conditional distribution $p(\y|\x)$. To do this, the generator and the discriminator are defined in terms of                                                                                                                                                                                                                                                     the condition variable $\x$ as well:
\begin{equation}\label{eqn:cgan}\small
\medmuskip=0mu
\thinmuskip=0mu
\thickmuskip=0mu
\min_{G}~\max_{D}\underbrace{\Big [ \mathbb{E}_{\x,\y} \log D(\x,~\y) + \mathbb{E}_{\x,\z} \log (1 - D ( \x,~ G(\x,~\z)) \Big ]}_{\L_{\text{cGAN}}(G, D)}.
\end{equation}
\texttt{pix2pix} \cite{isola2017pix2pix} is an image translation model that learns a mapping between two distinct domains of images, such as aerial photos and maps.
A \texttt{pix2pix} model takes paired images $(\x, \y)$ as its training data and minimizes the conditional GAN loss along with an additional L1 loss:
\begin{equation}\label{eqn:l1}\small
\mathbb{E}_{\mathbf{x},\mathbf{y},\mathbf{z}} ~ \big \lVert ~ \y - G(\x, \z) ~ \big \rVert_1,
\end{equation}
which encourages the conditional generator to learn the forward mapping from $\mathbf{x}$ to $\mathbf{y}$. It can be thought that the GAN loss in Equation \ref{eqn:cgan} is fine-tuning the mapping learned by the L1 loss in Equation \ref{eqn:l1}, resulting in a predictive mapping that better respects the probabilistic dependencies within the labels $\y$.

In this paper, we adapt this approach to music transcription tasks and show that we can indeed improve the performance by introducing an adversarial loss to an existing music transcription model.

\section{Method}

\begin{figure}[t]
	\includegraphics[width=\columnwidth]{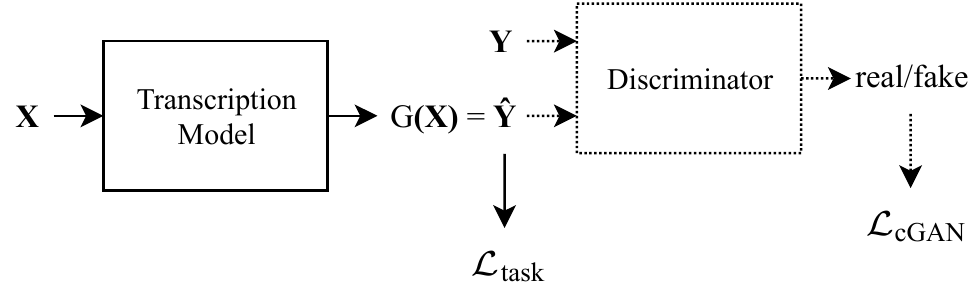}
	
	\caption{A computation graph showing how a discriminator is appended to the original model. The appended parts are shown as dotted components.}\label{fig:discriminator}
\end{figure}

We describe a general method for improving an NN-based transcription model $G$ that performs prediction of a two-dimensional target $\Y$ from an input audio representation $\X$.
Say the original model $G$ is trained by minimizing the loss $\L_{\textrm{task}}(G(\X), \Y)$ between the predicted target $\hat{\Y} = G(\X)$ and the ground-truth $\Y$. 
The main idea of our method is to adapt \texttt{pix2pix}~\cite{isola2017pix2pix} to this setup, by introducing an adversarial discriminator $D$ during the training process.
The adversarial training objective includes the conditional GAN loss $\L_{\text{cGAN}}$ (Equation \ref{eqn:cgan}):
\begin{equation}\small
\min_{G}\max_{D} \E_{\X,\Y} \Big [ \nu\L_{\text{task}} (G(\X), \Y) + \L_{\text{cGAN}}(G, D) \Big ],
\end{equation}
where $\nu$ is a hyperparameter that controls how much the conditional GAN loss contributes to the gradient steps relative to the discriminative loss $\L_{\text{task}}$.
Figure \ref{fig:discriminator} illustrates how the two components are connected in the computation graph and how the loss terms are calculated.

Adversarial training with $\L_{\text{cGAN}}$ allows the model to learn the inter-label dependencies as desired, even when $\L_{\text{task}}$ is defined only in terms of element-wise operations between $\hat{\Y}$ and $\Y$, as in Equation \ref{eqn:elementwise}.
In the next subsection, we describe a neural network architecture for the  cGAN discriminator that leverages prior knowledge on music.

\subsection{Musically Inspired Adversarial Discriminator}

Following \texttt{pix2pix}, we use a fully convolutional architecture~\cite{long2015fcn} for the discriminator.
By being fully convolutional, the discriminator has translation invariance not only along the time axis (as in HMMs and RNNs) but also along the frequency axis.
Since the discriminator determines how realistic a polyphonic note sequence is, the translation invariance enforces that the decision does not depend on the musical key, but only on the relative pitch and time intervals between the notes.
This effectively implements a music language model (MLM)~\cite{boulanger2012rnnnade,sigtia2016endtoend} and biases the transcription toward more realistic note sequences.

Unlike the image-to-image translation problem, the input representations (e.g.~Mel spectrograms) and the output representations (e.g.~piano rolls) of a music transcription model can have different dimensions.
This makes combining $\X$ and $\Y$ in a fully convolutional manner difficult.
For this reason, we make the discriminator a function of $\Y$ only, simplifying the objective in Equation \ref{eqn:cgan} to:
\begin{equation}\label{eqn:simple-cgan}\small
\L_{\text{cGAN}}(G, D) =  \mathbb{E}_{\Y}  \log D(\Y) + \mathbb{E}_{\X} \log (1 - D ( G(\X))).
\end{equation}
Note that $\z$ is also omitted in Equation \ref{eqn:simple-cgan}, as we follow \cite{isola2017pix2pix} and implement the stochasticity of $\z$ only in terms of dropout layers \cite{srivastava2014dropout}, without explicitly feeding random noises into the generator.
This causes a mode collapse problem where the learned $p(\Y|\X)$ is not diverse enough, but it does not harm our purpose of producing more realistic target representations.

\subsection{TTUR and \textit{mixup} to Stabilize GAN Training}

Although an ideal GAN generator can fully reconstruct the data distribution at the global optimum~\cite{goodfellow2014gan}, training of GANs in practice is notoriously difficult, especially for high-dimensional data~\cite{goodfellow2016gan}.
This led to the inventions of a plethora of techniques for stabilizing GAN training, among which we employ the two-timescale update rule (TTUR)~\cite{heusel2017ttur} and \textit{mixup}~\cite{zhang2018mixup}.
TTUR means simply setting the generator's learning rate a few times larger than that of the discriminator, which has been empirically shown to stabilize GAN training significantly.

The other technique, \textit{mixup}, is an extension to empirical risk minimization where training data samples are drawn from convex interpolations between pairs of empirical data samples.
For a pair of feature-target tuples $(\X_i, \Y_i)$ and $(\X_j, \Y_j)$ sampled randomly from the empirical distribution, their convex interpolation is given by:
\begin{equation}\small
\begin{array}{l@{}l}
\tilde{\X} = \lambda \X_i + (1 - \lambda) \X_j \\
\tilde{\Y} = \lambda \Y_i + (1 - \lambda) \Y_j
\end{array}
\end{equation}
where $\lambda \sim \text{Beta}(\alpha, \alpha)$, and $\alpha$ is the \textit{mixup} hyperparameter which controls the strength of interpolation.
When $\alpha = 0$, the Beta distribution becomes $\text{Bernoulli}(0.5)$ which recovers the usual GAN training without \textit{mixup}.

\setlength{\algomargin}{0em}
\DontPrintSemicolon
\begin{algorithm2e}[b!]
	\vspace{-0.3em}\noindent
	\rule{\columnwidth}{0.75pt}\;
	\label{alg:training}
	\caption{Training of a \textit{mixup} Conditional GAN.}
	\small\setstretch{1.05}
	\KwIn{
		Generator $G_\vartheta(\X)$ with initial parameters $\vartheta$, learning rate $\eta$, and loss function $\L_{\text{task}}(\hat{\Y}, \Y)$, discriminator $D_\varphi(\Y)$ with initial parameters $\varphi$, learning rate $\beta$, and loss function $\ell \in \{\text{BCE}, \text{MSE}\}$, batch size $m$, training data distribution $p(\X, \Y)$,  \texttt{pix2pix} weight $\nu$, \textit{mixup} strength $\alpha$.
	}
	\KwOut{Trained conditional generator $G_\vartheta(\X)$.}
	\vspace{-0.5em}\noindent
	\rule{\columnwidth}{0.5pt}\;
	\While{$\varphi$ and $\vartheta$ have not converged}{
		$\{(\X_i, \Y_i)\}_{i=1,\cdots,m} \leftarrow m \text{ samples from } p(\X, \Y)$\;
		\For{$i = 1, \cdots, m$}{
			$\hat{\Y}_i \leftarrow G_\vartheta(\X_i)$\;
			$\lambda_i \leftarrow \text{sample from Beta}(\alpha, \alpha)$\;
			$\tilde{\Y}_i \leftarrow \lambda_i \Y_i + (1 - \lambda_i) \hat{\Y}_i$
		}
		$\L_{\text{cGAN}}^D\>\leftarrow \textstyle\sum_{i=1}^M \ell(D_\varphi(\tilde{\Y}_i), \lambda_i)$\;
		$\varphi \leftarrow \varphi - \beta \cdot \nabla_\varphi \L_{\text{cGAN}}^D $\;
		$\L_{\text{cGAN}}^G \leftarrow \textstyle\sum_{i=1}^M \ell (D_\varphi(\tilde{\Y}_i), 1 - \lambda_i) $\;
		$\vartheta \leftarrow \vartheta - \eta \cdot \textstyle\nabla_\vartheta \Big [ \sum_{i=1}^m \nu \L_{\text{task}}(\hat{\Y}_i, Y_i) - \L_{\text{cGAN}}^G \Big ] $
	}
	\vspace{-0.5em}\noindent
	\rule{\columnwidth}{0.75pt}\;
\end{algorithm2e}

\textit{mixup} is readily applicable to the binary classification task of GAN discriminators.
In our conditional GAN setup, we have an additional advantage of having paired samples of a real label $\Y$ and a fake label $\hat{\Y} = G(\X)$, which allow us to replace Equation \ref{eqn:simple-cgan} with:
\begin{equation}\label{eqn:mixup-gan}\small
\min_{G} \max_{D} \mathbb{E}_{\X,\Y,\lambda} \Big [ - \ell (D(\lambda \Y + (1 - \lambda) G(\X) ), \lambda) \Big ].
\end{equation}
where $\ell(p, y) = - y \log p - (1-y) \log (1-p)$ is the binary cross entropy (BCE) function.
With this \textit{mixup} setup, the discriminator now has to operate on the convex interpolation between the predicted representation and the corresponding ground truth.
This makes the discriminator's task even more difficult when the prediction gets close to the ground truth, which is desirable because the discrimiantor should be inconclusive (i.e. $D = \tfrac{1}{2}$ everywhere) at the global optimum~\cite{goodfellow2014gan}.

Algorithm \ref{alg:training} details the procedure of training the conditional GAN using \textit{mixup}, based on Equations \ref{eqn:simple-cgan} and \ref{eqn:mixup-gan}.
Note that for training the generator network, we perform label flipping in $\L_{\text{cGAN}}^G$ similarly as in Equation \ref{eqn:nsgan}.
Also, to train a least-squares GAN (Equation \ref{eqn:lsgan}) instead, we can simply replace $\ell$ with a mean squared error (MSE) loss.

\begin{table}[b!]
	\small
	\renewcommand\arraystretch{1.2}
	\renewcommand{\tabcolsep}{5pt}
	\centering
	\begin{tabular}{l c} \toprule
		Hyperparameter & Values \\ \hline
		Generator learning rate $\eta$ & 0.0006 \\ 
		Discriminator learning rate $\beta$ & 0.0001 \\
		Discriminator loss function $\ell$ & \{BCE, MSE\}\\
		Batch size $m$ & 8 \\
		\texttt{pix2pix} weight $\nu$ & 100 \\
		\textit{mixup} strength $\alpha$ & \{0, 0.2, 0.3, 0.4\} \\
		Activation threshold $\tau$ & 0.5 \\
		Training sequence length & 327,680 \\
		\bottomrule
	\end{tabular}
	\caption{Hyperparameters used during the experiments.}\label{tab:hyperparameters}
\end{table}

\setcounter{table}{2}
\begin{table*}[!b]
	\centering
	\scriptsize
	\renewcommand\arraystretch{1.5}
	\renewcommand{\tabcolsep}{0em}
	\newcolumntype{M}[1]{>{\centering\arraybackslash}b{#1}}
	\newcolumntype{C}{>{\centering\arraybackslash}m{2.4em}}
	\begin{tabular}{@{\extracolsep{1em}}lM{5em}CCCCCCCCCCCCCCCC}
		&\multirow{2}{4em}{\centering\textit{mixup} strength $\alpha$} &\multicolumn{7}{M{23em}}{Frame Metrics}&\multicolumn{3}{M{9em}}{Note Metrics} &\multicolumn{3}{M{9em}}{Note Metrics with Offsets} &\multicolumn{3}{M{9em}}{Note Metrics with Offsets \& Velocity} \\ \cline{3-9} \cline{10-12} \cline{13-15} \cline{16-18}
		& & F1 & P & R & $E_\text{total}$ & $E_\text{subs}$ & $E_\text{miss}$ & $E_\text{fa}$ & F1 & P & R & F1 & P & R & F1 & P & R \\ \hline
		Baseline & & .899 & .946 & .857 & .179 & .013 & .130 & .036 & .942 & .990 & .899 & .802 & .842 & .765 & .790 & .830 & .755 \\
		Non-Saturating GAN & 0.3 & \textbf{.914} & .931 & .898 & \textbf{.156} & .012 & .089 & .054 & \textbf{.956} & .981 & .932 & \textbf{.813} & .835 & .793 & \textbf{.802} & .823 & .782 \\
		Least-Squares GAN & 0.3 & .906 & .942 & .875 & .167 & .013 & .113 & .042 & .950 & .988 & .916 & .810 & .841 & .781 & .799 & .830 & .771 \\
		\hline \end{tabular}
	\caption{Summary of transcription performance using \textit{mixup} strength $\alpha = 0.3$. The non-saturating GAN loss has the highest performance across all F1 metrics. The average metrics across the tracks in the MAESTRO test dataset are reported, and the model checkpoint where the average of frame F1 and note F1 is the highest on the validation dataset is used.}\label{tab:performance}
\end{table*}

\section{Experimental Setup}

To verify the effectiveness of our approach, we compare Onsets and Frames~\cite{hawthorne2018onsetsframes}, a state-of-the-art piano transcription model, with variants of the same model that are trained with the adversarial loss.
We also aim to evaluate the choices of the GAN loss and the \textit{mixup} strength $\alpha$.

\subsection{Model Architecture}

We use the extended Onsets and Frames model~\cite{hawthorne2019maestro} which increased the CNN channels to 48/48/96, the LSTM units to 256, and the FC units to 768.
The extended model has total 26.5 million parameters.
We do not use the frame loss weights described in~\cite{hawthorne2018onsetsframes} in favor of the offset stack introduced in the extended version (see Figure~\ref{fig:onsetsframes}).
During inference, we first calculate the posteriors corresponding to overlapping chunks of audio, with the same length as the training sequences, and perform overlap-add using Hamming windows to obtain the full-length posterior.
This is because the effects of adversarial learning do not continue further than the training sequence length when we let the recurrent networks continue to predict longer sequences.

The input to the discriminator has two channels for the onset and frame predictions.
The discriminator has 5 convolutional layers: \texttt{c32k3s2p1}, \texttt{c64k3s2p1}, \texttt{c128k3s2p1}, \texttt{c256k3s2p1}, \texttt{c1k5s1p2}, where the numbers indicate the number of output channels, the kernel size, the stride amount, and the padding size.
At each non-final layer, dropout of probability 0.5 and leaky ReLU activation with negative slope 0.2 are used.
The mean of the final layer output along the time and frequency axes is taken as the discriminator output.

\subsection{Hyperparameters}

Table \ref{tab:hyperparameters} summarizes the hyperparameters used during the experiments, which are mostly taken directly from~\cite{hawthorne2018onsetsframes} and \cite{isola2017pix2pix}.
Also following~\cite{hawthorne2018onsetsframes}, we use Adam~\cite{kingma2015adam} and apply learning rate decay of factor 0.98 in every 10,000 iterations, for both the generator and the discriminator.
We examine two types of GAN losses, the non-saturating GAN ($\ell = \text{BCE}$) and the least-squares GAN ($\ell = \text{MSE}$).
For each GAN loss, multiple values of \textit{mixup} strengths are compared with $\alpha = 0$, i.e. no \textit{mixup}.
Training runs for one million iterations, and the iteration that best performs on the validation set are used for evaluation on the test set.

\subsection{Dataset}

We use the MAESTRO dataset~\cite{hawthorne2019maestro}, which contains Disklavier recordings of 1,184 classical piano performances.
The dataset consists of 172.3 hours of audio, which are provided with 140.1, 15.3, and 16.9 hours of train/validation/test splits such that recordings of one composition only appear in the same split.
We resample the audio to 16 kHz and down-mix into a single channel.
Following~\cite{hawthorne2018onsetsframes}, an STFT window of 2,048 samples is used for producing 229-bin Mel spectrograms, and a hop length of 32 ms is used.
Training sequences sliced at random positions are used, unlike the official implementation which slices training sequences at silence or zero crossings.

\subsection{Evaluation Metrics}

The Onsets and Frames model perform both frame-level and note-level predictions, and their performance can be evaluated with the standard precision, recall, and F1 metrics.
For multi-pitch estimation, we also report the error rate metrics defined in~\cite{poliner2006discriminative}, which include total error, substitution error, miss error, and false alarm error.
We use the \texttt{mir\_eval}~\cite{raffel2014mireval} library for all metric calculations.
For the note-level metrics, we use the default settings of the library, which use 50 ms for the onset tolerance, 50 ms or 20\% of the note length (whichever is longer) for the offset tolerance, and 0.1 for the velocity tolerance.

\section{Results}


\subsection{Comparison with the Baseline Metrics}

\begin{figure*}[t]
\centering
\minipage{1.03\textwidth}
\hspace{-0.02\textwidth}
\includegraphics[width=0.333\textwidth]{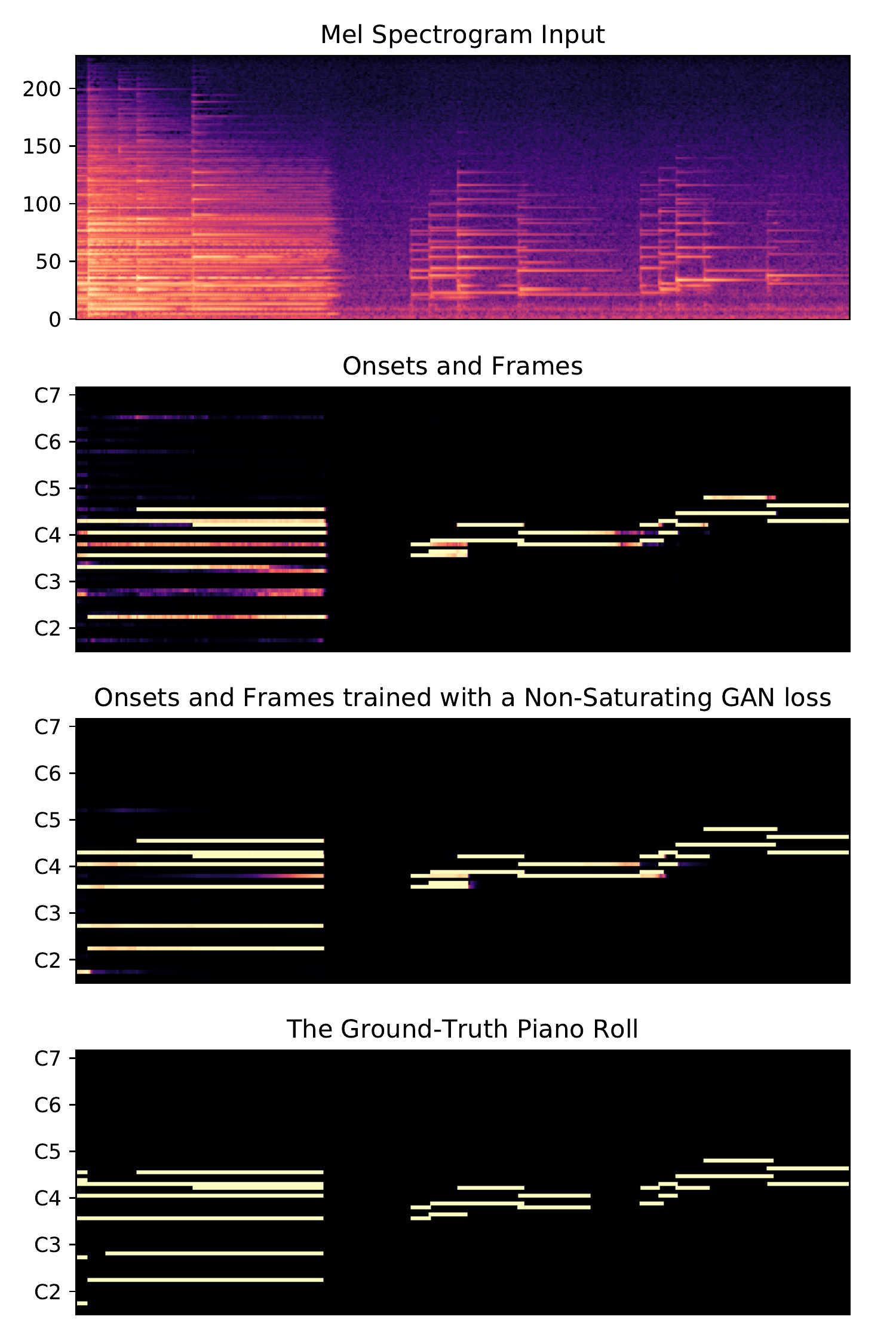}%
\includegraphics[width=0.333\textwidth]{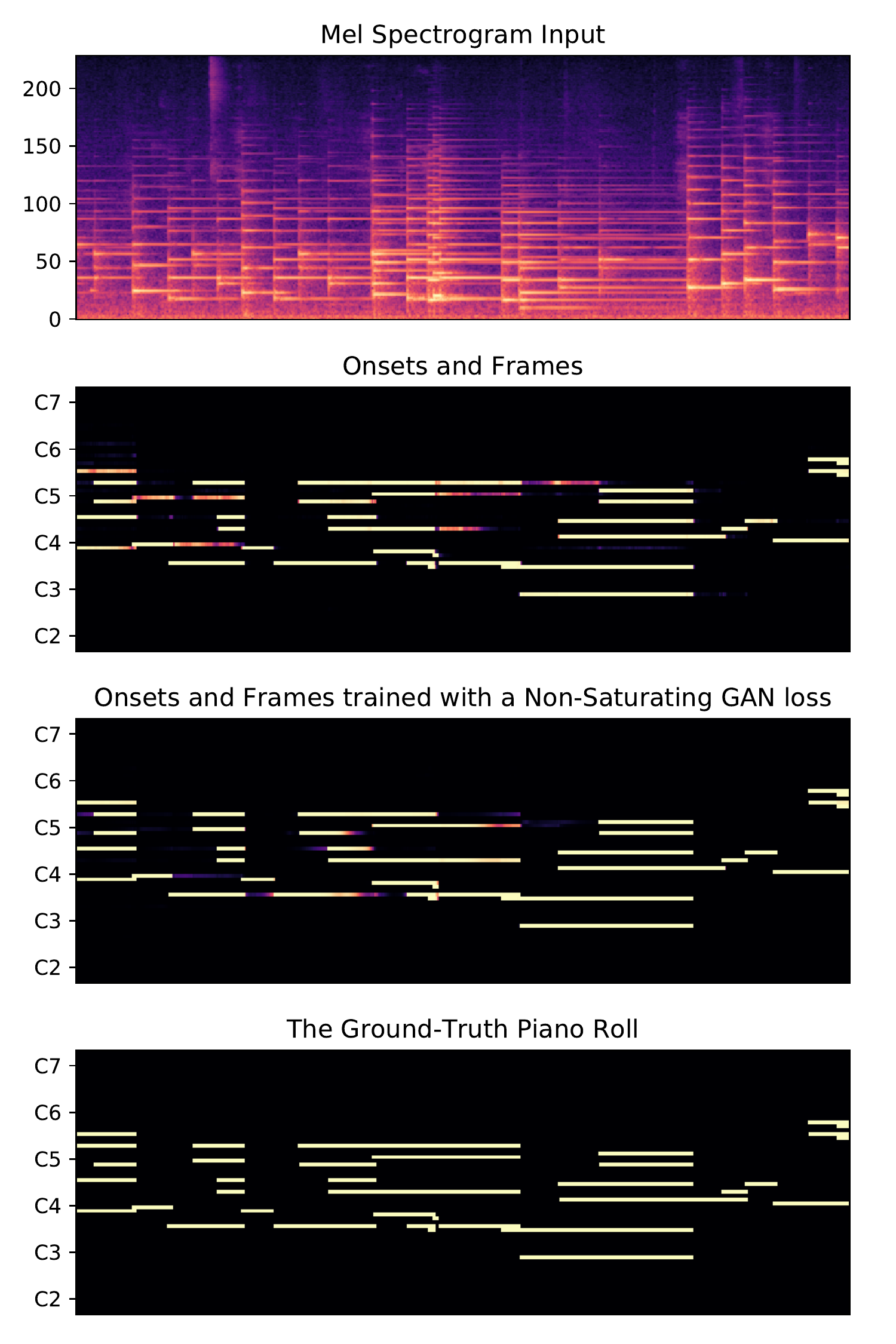}%
\includegraphics[width=0.333\textwidth]{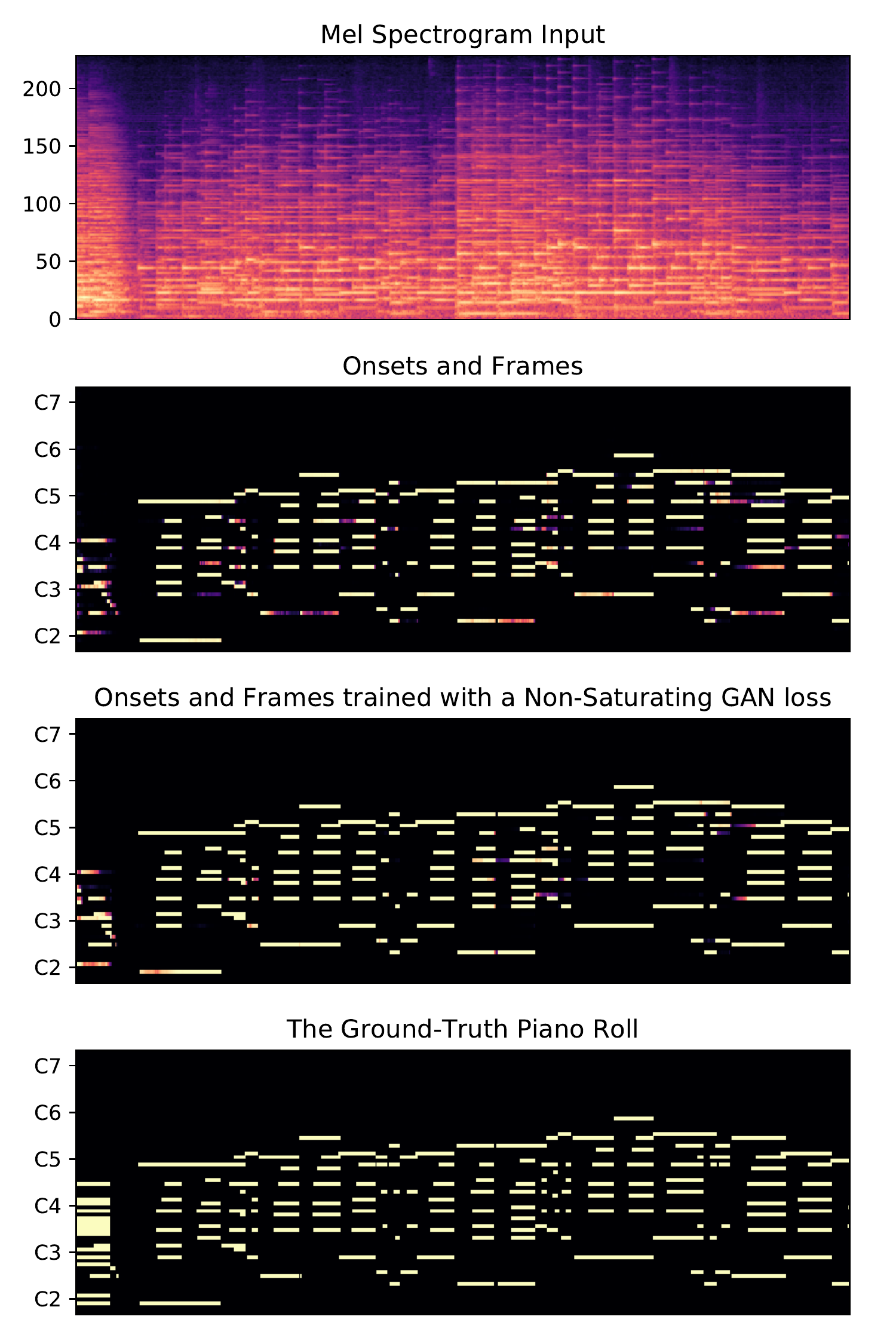}%
\endminipage
\caption{Comparisons of the frame activation posterior predicted by the baseline and our model ($\ell = \text{BCE}$, $\alpha = 0.3$), on three example segments. The input Mel spectrograms and the target piano rolls are shown together. The GAN version produces more confident predictions compared to the noisy baselines, leading to more accurate predictions.}\label{fig:predictions}
\end{figure*}

\begin{figure*}[t]
	\minipage{0.37\textwidth}
		\includegraphics[height=9.2em]{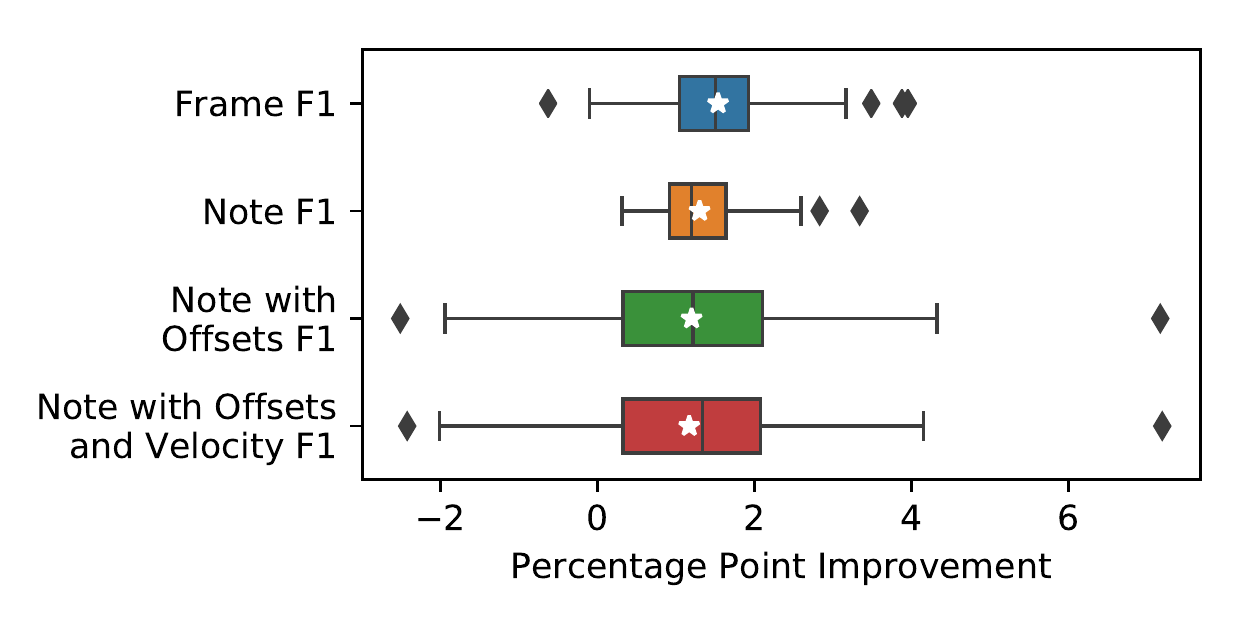}
		\caption{F1 score improvements over the baseline, tested on the MAESTRO test tracks.}\label{fig:pertrack}
	\endminipage\hfill
	\minipage{0.22\textwidth}
		\includegraphics[height=9.2em]{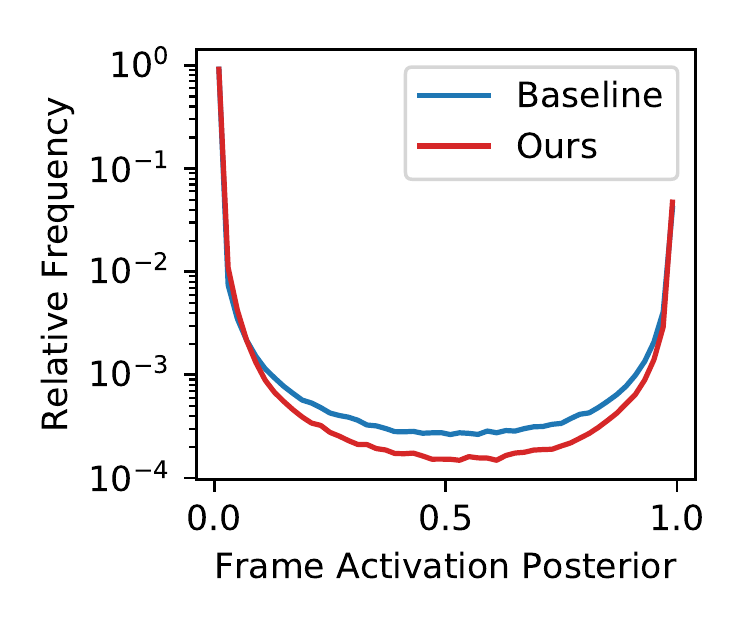}
		\caption{Distribution of frame activation values.}\label{fig:distribution}
	\endminipage\hfill
	\minipage{0.37\textwidth}
		\includegraphics[height=9.2em]{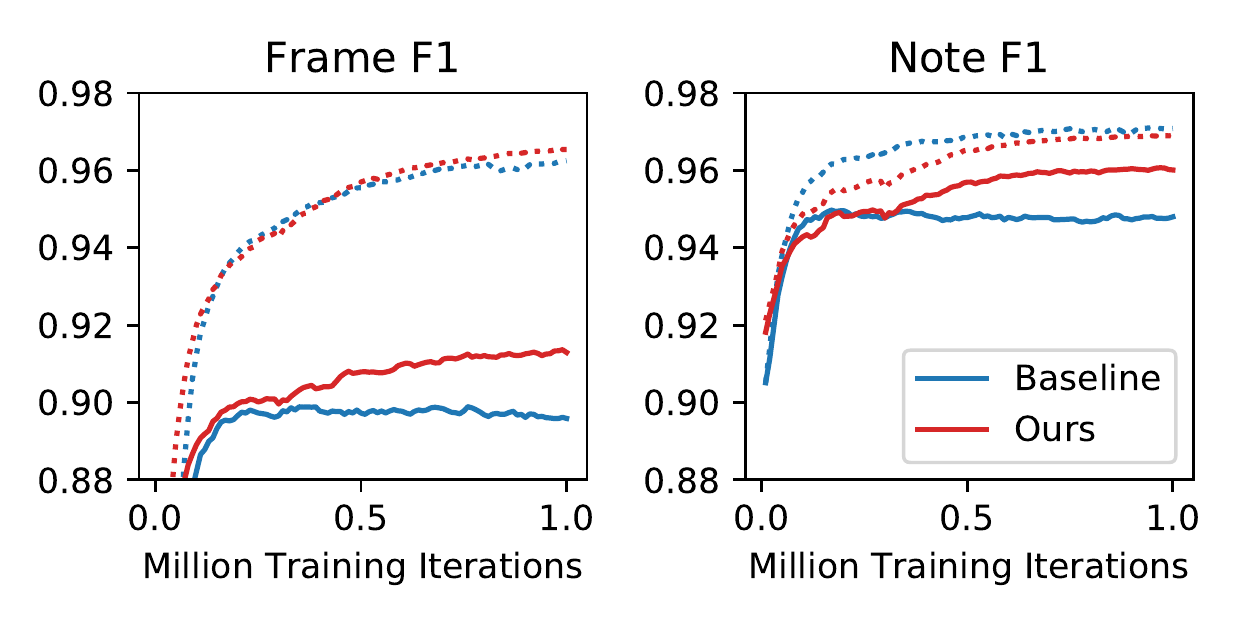}
		\caption{Learning curves showing the generalization gaps; training curves are dotted.}\label{fig:training}
	\endminipage
\end{figure*}

Table \ref{tab:alpha} and \ref{tab:performance} summarize the transcription performance, clearly showing a consistent improvement in the conditional GAN models over the Onsets and Frames baseline.
Table \ref{tab:alpha} shows that both non-saturating GAN and least-squares GAN achieve the highest frame and note F1 scores when the \textit{mixup} strength $\alpha = 0.3$ is used, and they both outperform the baseline.
The binary piano rolls are easy to distinguish from the non-binary predictions, which may cause imbalanced adversarial training. \textit{mixup} allows non-binary piano rolls to be fed to the discriminator, making its task more challenging and leading to higher performance.

Table \ref{tab:performance} shows an important trend of the cGAN results compared to the baseline that cGAN trades off a bit of precision for a significant improvement in recall; this is a side effect of the cGAN producing more confident predictions, as will be discussed in the following subsections.

While the percentage differences are moderate, our method achieves statistically significant improvements in F1 metrics on the MAESTRO test dataset ($p < 10^{-14}$ for all 4 metrics, two-tailed paired $t$-test).
The distribution of per-track improvement in each F1 metric is shown in Figure \ref{fig:pertrack}, which indicates that the improvements are evenly distributed across the majority of the tracks.
These improvements are especially promising, considering that Onsets and Frames is already a very strong baseline.

\subsection{Visualization of Frame Activations}

To better understand the inner workings of the conditional GAN framework, we visualize the frame posteriorgrams created by the baseline and the best performing conditional GAN model in Figure \ref{fig:predictions}.
In contrast to the baseline posteriorgrams which have many blurry segments, the posteriorgrams generated by our method mostly contain segments with solid colors, meaning that the model is more confident in its prediction.
Figure~\ref{fig:distribution} shows that the proportion of frame activation values in $(0.1, 0.9)$ is noticeably higher in the baseline, thus making the output less sensitive to the threshold choice.
This is because indecisive predictions are penalized by the discriminator, since they are easy to distinguish from the ground-truth which contains only binary labels.
The generator is therefore encouraged to output the most probable note sequences even when it is unsure, rather than producing blurry posteriorgrams that might hamper the decoding process.
This allows for an interpretation in which the GAN loss provides a prior for valid onset and frame activations, and the model learns to perform MAP estimation based on this prior.

\setcounter{table}{1}
\begin{table}[t]
	\centering
	\scriptsize
	\renewcommand\arraystretch{1.5}
	\renewcommand{\tabcolsep}{0em}
	\newcolumntype{M}[1]{>{\centering\arraybackslash}b{#1}}
	\newcolumntype{C}{>{\centering\arraybackslash}m{2.4em}}
	\begin{tabular}{@{\extracolsep{1em}}lM{4em}M{8em}CCCC}
		& & & \multicolumn{4}{c}{\textit{mixup} strength $\alpha$} \\ \cline{4-7}
		& Baseline & GAN type & 0 & 0.2 & 0.3 & 0.4 \\ \hline
		\multirow{2.25}{4em}{Frame F1} & \multirow{2.25}{4em}{\centering0.899} & Non-Saturating & 0.664 & 0.912 & \textbf{0.914} & 0.907 \\
		& & Least-Squares & 0.904 & 0.903 & 0.906 & 0.898 \\ \hline
		\multirow{2.25}{4em}{Note F1} & \multirow{2.25}{4em}{\centering0.942} & Non-Saturating & 0.717 & 0.953 & \textbf{0.956} & 0.951 \\
		& & Least-Squares & 0.944 & 0.947 & 0.950 & 0.943 \\
	\hline
	\end{tabular}
	\caption{Frame and note F1 scores are the highest when the non-saturating GAN loss and $\alpha = 0.3$ are used.}\label{tab:alpha}
\end{table}

\subsection{Training Dynamics and The Generalization Gap}

Figure \ref{fig:training} shows the learning curves for the frame F1 and note F1 scores, where the scores on the training dataset are plotted in dotted lines.
It is noticeable in the figure that the validation F1 scores for the baseline stagnate after 300k iterations, while the F1 scores of our model steadily grow until the end of 1 million iterations.
Thanks to this, the generalization gap --- the difference between the training and validation F1 scores --- is significantly smaller for the conditional GAN model.
This means that the GAN loss works as an effective regularizer that encourages the trained model to generalize better to unseen data, rather than memorizing the note sequences in the training dataset as LSTMs are known to be capable of~\cite{zaremba2014recurrent}.

\section{Conclusions}

We have presented an adversarial training method that can consistently outperform the baseline Onsets and Frames model, using the standard frame-level and note-level transcription metrics and visualizations that show how the improved model predicts more confident output.
To achieve this, a discriminator network is trained competitively with the transcription model, i.e. a conditional generator, so that the discriminator serves as a learned regularizer that provides a prior for realistic note sequences.

Our results show that modeling the inter-label dependencies in the target distribution is important and brings measurable performance improvements.
Our method is generic, and any model that involves predicting two-dimensional representation should be able to benefit from including an adversarial loss.
These approaches are common not only in transcription models but also in speech or music synthesis models that predict spectrograms as an intermediate representation~\cite{shen2018tacotron2,kim2019mel2mel}.

Our results do not include the effects of using data augmentation~\cite{hawthorne2019maestro}, which is orthogonal to our approach and should bring additional performance improvements when applied.
As discussed, the discriminator imposes the prior on the target domain whereas data augmentation enriches the input audio distribution.
This implies that our method would be less effective when the majority of errors are due to the discrepancy in the audio distribution between the training and test datasets.
How to apply adversarial learning for better generalization on the input distribution is a potential future research direction.

\setstretch{0.92}
\bibliography{adversarial}

\begin{thebibliography}{10}

\bibitem{abdallah2006sparse}
Samer~A Abdallah and Mark~D Plumbley.
\newblock Unsupervised analysis of polyphonic music by sparse coding.
\newblock {\em IEEE Transactions on Neural Networks}, 17(1):179--196, 2006.

\bibitem{benetos2013multiple}
Emmanouil Benetos and Simon Dixon.
\newblock Multiple-instrument polyphonic music transcription using a temporally
  constrained shift-invariant model.
\newblock {\em The Journal of the Acoustical Society of America},
  133(3):1727--1741, 2013.

\bibitem{benetos2019amt}
Emmanouil Benetos, Simon Dixon, Zhiyao Duan, and Sebastian Ewert.
\newblock Automatic music transcription: An overview.
\newblock {\em IEEE Signal Processing Magazine}, 36(1):20--30, 2019.

\bibitem{benetos2013amt}
Emmanouil Benetos, Simon Dixon, Dimitrios Giannoulis, Holger Kirchhoff, and
  Anssi Klapuri.
\newblock Automatic music transcription: challenges and future directions.
\newblock {\em Journal of Intelligent Information Systems}, 41(3):407--434,
  2013.

\bibitem{bittner2017deepsalience}
Rachel~M Bittner, Brian McFee, Justin Salamon, Peter Li, and Juan~Pablo Bello.
\newblock Deep salience representations for f0 estimation in polyphonic music.
\newblock In {\em Proceedings of the International Society for Music
  Information Retrieval {(ISMIR)} Conference}, pages 63--70, 2017.

\bibitem{bock2012rnn}
Sebastian B{\"o}ck and Markus Schedl.
\newblock Polyphonic piano note transcription with recurrent neural networks.
\newblock In {\em Proceedings of the IEEE international conference on
  acoustics, speech and signal processing (ICASSP)}, pages 121--124, 2012.

\bibitem{boulanger2012rnnnade}
Nicolas Boulanger{-}Lewandowski, Yoshua Bengio, and Pascal Vincent.
\newblock Modeling temporal dependencies in high-dimensional sequences:
  Application to polyphonic music generation and transcription.
\newblock In {\em Proceedings of the International Conference on Machine
  Learning {(ICML)}}, 2012.

\bibitem{creswell2018gan}
Antonia Creswell, Tom White, Vincent Dumoulin, Kai Arulkumaran, Biswa Sengupta,
  and Anil~A Bharath.
\newblock Generative adversarial networks: An overview.
\newblock {\em IEEE Signal Processing Magazine}, 35(1):53--65, 2018.

\bibitem{dong2018musegan}
Hao-Wen Dong, Wen-Yi Hsiao, Li-Chia Yang, and Yi-Hsuan Yang.
\newblock Musegan: Multi-track sequential generative adversarial networks for
  symbolic music generation and accompaniment.
\newblock In {\em Thirty-Second AAAI Conference on Artificial Intelligence},
  2018.

\bibitem{dosovitskiy2016generating}
Alexey Dosovitskiy and Thomas Brox.
\newblock Generating images with perceptual similarity metrics based on deep
  networks.
\newblock In {\em Advances in Neural Information Processing Systems}, pages
  658--666, 2016.

\bibitem{engel2019gansynth}
Jesse Engel, Kumar~Krishna Agrawal, Shuo Chen, Ishaan Gulrajani, Chris Donahue,
  and Adam Roberts.
\newblock {GANSynth}: Adversarial neural audio synthesis.
\newblock {\em arXiv preprint arXiv:1902.08710}, 2019.

\bibitem{ewert2016admm}
Sebastian Ewert and Mark Sandler.
\newblock Piano transcription in the studio using an extensible alternating
  directions framework.
\newblock {\em IEEE/ACM Transactions on Audio, Speech, and Language
  Processing}, 24(11):1983--1997, 2016.

\bibitem{fevotte2011betanmf}
C{\'e}dric F{\'e}votte and J{\'e}r{\^o}me Idier.
\newblock Algorithms for nonnegative matrix factorization with the
  $\beta$-divergence.
\newblock {\em Neural computation}, 23(9):2421--2456, 2011.

\bibitem{fuentes2013harmonic}
Benoit Fuentes, Roland Badeau, and Ga{\"e}l Richard.
\newblock Harmonic adaptive latent component analysis of audio and application
  to music transcription.
\newblock {\em IEEE Transactions on Audio, Speech, and Language Processing},
  21(9):1854--1866, 2013.

\bibitem{goodfellow2016gan}
Ian Goodfellow.
\newblock {NIPS} 2016 tutorial: Generative adversarial networks.
\newblock {\em arXiv preprint arXiv:1701.00160}, 2016.

\bibitem{goodfellow2014gan}
Ian Goodfellow, Jean Pouget-Abadie, Mehdi Mirza, Bing Xu, David Warde-Farley,
  Sherjil Ozair, Aaron Courville, and Yoshua Bengio.
\newblock Generative adversarial nets.
\newblock In {\em Advances in Neural Information Processing Systems}, pages
  2672--2680, 2014.

\bibitem{hawthorne2018onsetsframes}
Curtis Hawthorne, Erich Elsen, Jialin Song, Adam Roberts, Ian Simon, Colin
  Raffel, Jesse Engel, Sageev Oore, and Douglas Eck.
\newblock Onsets and frames: Dual-objective piano transcription.
\newblock In {\em Proceedings of the International Society for Music
  Information Retrieval {(ISMIR)} Conference}, pages 50--57, 2018.

\bibitem{hawthorne2019maestro}
Curtis Hawthorne, Andrew Stasyuk, Adam Roberts, Ian Simon, Cheng-Zhi~Anna
  Huang, Sander Dieleman, Erich Elsen, Jesse Engel, and Douglas Eck.
\newblock Enabling factorized piano music modeling and generation with the
  {MAESTRO} dataset.
\newblock In {\em Proceedings of the International Conference on Learning
  Representations {(ICLR)}}, 2019.

\bibitem{heusel2017ttur}
Martin Heusel, Hubert Ramsauer, Thomas Unterthiner, Bernhard Nessler, and Sepp
  Hochreiter.
\newblock Gans trained by a two time-scale update rule converge to a local nash
  equilibrium.
\newblock In {\em Advances in Neural Information Processing Systems}, pages
  6626--6637, 2017.

\bibitem{hinton2006dbn}
Geoffrey~E Hinton, Simon Osindero, and Yee-Whye Teh.
\newblock A fast learning algorithm for deep belief nets.
\newblock {\em Neural Computation}, 18(7):1527--1554, 2006.

\bibitem{isola2017pix2pix}
Phillip Isola, Jun-Yan Zhu, Tinghui Zhou, and Alexei~A Efros.
\newblock Image-to-image translation with conditional adversarial networks.
\newblock In {\em Proceedings of the IEEE Conference on Computer Vision and
  Pattern Recognition {(CVPR)}}, pages 1125--1134, 2017.

\bibitem{karras2018stylegan}
Tero Karras, Samuli Laine, and Timo Aila.
\newblock A style-based generator architecture for generative adversarial
  networks.
\newblock {\em arXiv preprint arXiv:1812.04948}, 2018.

\bibitem{kelz2016framewise}
Rainer Kelz, Matthias Dorfer, Filip Korzeniowski, Sebastian B{\"{o}}ck, Andreas
  Arzt, and Gerhard Widmer.
\newblock On the potential of simple framewise approaches to piano
  transcription.
\newblock In {\em Proceedings of the International Society for Music
  Information Retrieval {(ISMIR)} Conference}, pages 475--481, 2016.

\bibitem{kim2019mel2mel}
Jong~Wook Kim, Rachel Bittner, Aparna Kumar, and Juan~Pablo Bello.
\newblock Neural music synthesis for flexible timbre control.
\newblock In {\em Proceedings of the {IEEE} International Conference on
  Acoustics, Speech and Signal Processing (ICASSP)}, 2019.

\bibitem{kim2018crepe}
Jong~Wook Kim, Justin Salamon, Peter Li, and Juan~Pablo Bello.
\newblock {CREPE}: A convolutional representation for pitch estimation.
\newblock In {\em Proceedings of the {IEEE} International Conference on
  Acoustics, Speech and Signal Processing (ICASSP)}, pages 161--165, 2018.

\bibitem{kingma2015adam}
Diederik~P. Kingma and Jimmy Ba.
\newblock Adam: {A} method for stochastic optimization.
\newblock In {\em Proceedings of the International Conference on Learning
  Representations, {(ICLR)}}, 2015.

\bibitem{larochelle2011nade}
Hugo Larochelle and Iain Murray.
\newblock The neural autoregressive distribution estimator.
\newblock In {\em Proceedings of the Fourteenth International Conference on
  Artificial Intelligence and Statistics}, pages 29--37, 2011.

\bibitem{lecun2015deeplearning}
Yann LeCun, Yoshua Bengio, and Geoffrey Hinton.
\newblock Deep learning.
\newblock {\em Nature}, 521(7553):436, 2015.

\bibitem{lee2001nmf}
Daniel~D Lee and H~Sebastian Seung.
\newblock Algorithms for non-negative matrix factorization.
\newblock In {\em Advances in Neural Information Processing Systems}, pages
  556--562, 2001.

\bibitem{long2015fcn}
Jonathan Long, Evan Shelhamer, and Trevor Darrell.
\newblock Fully convolutional networks for semantic segmentation.
\newblock In {\em Proceedings of the IEEE Conference on Computer Vision and
  Pattern Recognition {(CVPR)}}, pages 3431--3440, 2015.

\bibitem{mao2017lsgan}
Xudong Mao, Qing Li, Haoran Xie, Raymond~YK Lau, Zhen Wang, and Stephen
  Paul~Smolley.
\newblock Least squares generative adversarial networks.
\newblock In {\em Proceedings of the IEEE International Conference on Computer
  Vision}, pages 2794--2802, 2017.

\bibitem{mauch2014pyin}
Matthias Mauch and Simon Dixon.
\newblock {pYIN}: A fundamental frequency estimator using probabilistic
  threshold distributions.
\newblock In {\em Proceedings of the IEEE International Conference on
  Acoustics, Speech and Signal Processing (ICASSP)}, pages 659--663. IEEE,
  2014.

\bibitem{mirza2014cgan}
Mehdi Mirza and Simon Osindero.
\newblock Conditional generative adversarial nets.
\newblock {\em arXiv preprint arXiv:1411.1784}, 2014.

\bibitem{nam2011classification}
Juhan Nam, Jiquan Ngiam, Honglak Lee, and Malcolm Slaney.
\newblock A classification-based polyphonic piano transcription approach using
  learned feature representations.
\newblock In {\em Proceedings of the 12th International Society for Music
  Information Retrieval {(ISMIR)} Conference}, pages 175--180, 2011.

\bibitem{ni2012hmm}
Yizhao Ni, Matt McVicar, Raul Santos-Rodriguez, and Tijl De~Bie.
\newblock An end-to-end machine learning system for harmonic analysis of music.
\newblock {\em IEEE Transactions on Audio, Speech, and Language Processing},
  20(6):1771--1783, 2012.

\bibitem{poliner2006discriminative}
Graham~E Poliner and Daniel~PW Ellis.
\newblock A discriminative model for polyphonic piano transcription.
\newblock {\em EURASIP Journal on Advances in Signal Processing},
  2007(1):048317, 2006.

\bibitem{raffel2014mireval}
Colin Raffel, Brian McFee, Eric~J Humphrey, Justin Salamon, Oriol Nieto, Dawen
  Liang, Daniel~PW Ellis, and C~Colin Raffel.
\newblock mir\_eval: A transparent implementation of common {MIR} metrics.
\newblock In {\em Proceedings of the International Society for Music
  Information Retrieval {(ISMIR)} Conference}, 2014.

\bibitem{shen2018tacotron2}
Jonathan Shen, Ruoming Pang, Ron~J Weiss, Mike Schuster, Navdeep Jaitly,
  Zongheng Yang, Zhifeng Chen, Yu~Zhang, Yuxuan Wang, Rj~Skerrv-Ryan, et~al.
\newblock Natural {TTS} synthesis by conditioning {WaveNet} on {Mel}
  spectrogram predictions.
\newblock In {\em Proceedings of the IEEE International Conference on
  Acoustics, Speech and Signal Processing (ICASSP)}, pages 4779--4783. IEEE,
  2018.

\bibitem{sigtia2016endtoend}
Siddharth Sigtia, Emmanouil Benetos, and Simon Dixon.
\newblock An end-to-end neural network for polyphonic piano music
  transcription.
\newblock {\em IEEE/ACM Transactions on Audio, Speech, and Language
  Processing}, 24(5):927--939, 2016.

\bibitem{smaragdis2003nmf}
Paris Smaragdis and Judith~C Brown.
\newblock Non-negative matrix factorization for polyphonic music transcription.
\newblock In {\em 2003 {IEEE} Workshop on Applications of Signal Processing to
  Audio and Acoustics}, pages 177--180, 2003.

\bibitem{srivastava2014dropout}
Nitish Srivastava, Geoffrey Hinton, Alex Krizhevsky, Ilya Sutskever, and Ruslan
  Salakhutdinov.
\newblock Dropout: a simple way to prevent neural networks from overfitting.
\newblock {\em The Journal of Machine Learning Research}, 15(1):1929--1958,
  2014.

\bibitem{van2016pixelcnn}
Aaron Van~den Oord, Nal Kalchbrenner, Lasse Espeholt, Oriol Vinyals, Alex
  Graves, et~al.
\newblock Conditional image generation with {PixelCNN} decoders.
\newblock In {\em Advances in Neural Information Processing Systems}, pages
  4790--4798, 2016.

\bibitem{van2016pixelrnn}
A{\"a}ron Van Den~Oord, Nal Kalchbrenner, and Koray Kavukcuoglu.
\newblock Pixel recurrent neural networks.
\newblock In {\em Proceedings of the International Conference on Machine
  Learning {(ICML)}}, pages 1747--1756, 2016.

\bibitem{vincent2010adaptive}
Emmanuel Vincent, Nancy Bertin, and Roland Badeau.
\newblock Adaptive harmonic spectral decomposition for multiple pitch
  estimation.
\newblock {\em IEEE Transactions on Audio, Speech, and Language Processing},
  18(3):528--537, 2010.

\bibitem{yang2017midinet}
Li{-}Chia Yang, Szu{-}Yu Chou, and Yi{-}Hsuan Yang.
\newblock Midinet: {A} convolutional generative adversarial network for
  symbolic-domain music generation.
\newblock In {\em Proceedings of the International Society for Music
  Information Retrieval {(ISMIR)} Conference}, pages 324--331, 2017.

\bibitem{zaremba2014recurrent}
Wojciech Zaremba, Ilya Sutskever, and Oriol Vinyals.
\newblock Recurrent neural network regularization.
\newblock {\em arXiv preprint arXiv:1409.2329}, 2014.

\bibitem{zhang2018mixup}
Hongyi Zhang, Moustapha Cisse, Yann~N. Dauphin, and David Lopez-Paz.
\newblock \textit{mixup}: Beyond empirical risk minimization.
\newblock In {\em Proceedings of the International Conference on Learning
  Representations {(ICLR)}}, 2018.

\end{thebibliography}

\end{document}